# A NOTE ON SEPARABLE NONLINEAR LEAST SQUARES PROBLEM

Wajeb Gharibi
Computer Science & Information Systems College,
Jazan University,
Jazan, KSA
Gharibiw2002@yahoo.com

Omar Saeed Al-Mushayt
Computer Science & Information Systems College,
Jazan University,
Jazan, KSA
oalmushayt@yahoo.com

*Abstract*—Separable nonlinear least squares (SNLS) problem is a special class of nonlinear least squares (NLS) problems, whose objective function is a mixture of linear and nonlinear functions. It has many applications in many different areas, especially in Operations Research and Computer Sciences. They are difficult to solve with the infinite-norm metric. In this paper, we give a short note on the separable nonlinear least squares problem, unseparated scheme for NLS, and propose an algorithm for solving mixed linear-nonlinear minimization problem, method of which results in solving a series of least squares separable problems.

## I. INTRODUCTION

Separable nonlinear least squares (SNLS) problem is a special class of nonlinear least squares (NLS) problems, whose objective function is a mixture of linear and nonlinear functions. It has many applications in many areas such as, Numerical Analysis, Mechanical Systems, Neural Networks, Telecommunications, Robotics and Environmental Sciences, to name just a few [1, 2].

The existing special algorithms for these problems are derived from the variable projections scheme proposed by Golub and Pereyra [1]. However, when the linear part of variables has some bounded constraints, the methods based on variable projection strategy will be invalid. Here, we propose an unseparated scheme for NLS and an algorithm which results in solving a series of least squares separable problems.

Given a set of observation $\{y_i\}$, a separable nonlinear squares problem can be defined as follows:

$$r(a,\alpha) = y_i - \sum_{j=1}^{n} a_j \phi_j(\alpha, t_i) \qquad (1)$$

where $t_i$ are independent variables associated with the observation $\{y_i\}$, while the $a_j$, and the $k$-dimensional vector $\alpha$ are the parameters to be determined by minimizing the LS functional $r(a,\alpha)$. We can write the above equation in the following matrix form:

$$\|r(a,\alpha)\|_2^2 = \|y - \Phi(\alpha)a\|_2^2, \qquad (2)$$

where the columns of matrix $\Phi$ correspond to the nonlinear functions $\phi_j(\alpha, t_i)$ of the $k$ parameters $\alpha$ evaluated at all the $t_i$ values, and the vectors **a** and **y** represent the linear parameters and the observations respectively.

It is easy to see that if we knew the nonlinear parameters $\alpha$, then the linear parameters **a** could be obtained by solving the linear least squares problem:

$$a = \Phi(\alpha)^+ y,$$

which stands for the minimum-norm solution of the linear least squares problem for fixed $\alpha$, where $\Phi(\alpha)^+$ is the Moor-Penrose generalized inverse of $\Phi(\alpha)$. By replacing this **a** in the original functional, we obtain:

$$\min_{\alpha} \frac{1}{2}\|r_2(\alpha)\|_2^2 = \min_{\alpha} \frac{1}{2}\|(I - \Phi(\alpha)\Phi(\alpha)^+)y\|^2, \qquad (3)$$

This is the Variable Projection functional which its name stems from the fact that the matrix in parenthesis is the projector on the orthogonal complement of the column space of $\Phi$ [1].

The rest of our paper is organized as follows. In the next section we present unseparated scheme for the NLS problems. In section 3, we propose our method; Lagrangian Dual Algorithm. Discussion is made in the last section.

## II. UNSEPARATED SCHEME FOR THE NLS PROBLEMS

Consider the following NLS problem:

$$\min_{x \in R^n} F(x) = \frac{1}{2} \|f(x)\|_2^2, \quad (4)$$

where $f(x) \in R^m$, with $(f(x))_i = f_i(x)$. Many kinds of iterative methods have already been designed to solve it. Most methods for NLS are based on the linear approximation of $f$ in each iteration, which is derived from Gauss-Newton method. We can describe the main idea of Gauss-Newton method as follows.

Suppose our current iterative point is $x_k$, then we obtain the next point $x_{k+1} = x_k + d_k$ by solving the following linear least square (LLS) problem:

$$\min_{d_k \in R^n} \frac{1}{2} \|f(x_k) + J(x_k)d_k\|_2^2. \quad (5)$$

Here $J(x)$ is the Jacobian of $f(x)$. We can get

$$d_k = -(J(x_k))^+ f(x_k) = -(J(x_k)^T J(x_k))^{-1} J(x_k) f(x_k). \quad (6)$$

If we compare (6) with the Newton step of (4), we can find that Gauss-Newton method uses $J(x_k)^T J(x_k)$, only containing the first order information of $f$, to substitute the real Hessian of $F(x)$

$$\nabla^2 F(x) = J(x)^T J(x) + \sum_{i=1}^m \nabla^2 f_i(x) f_i(x), \quad (7)$$

by omitting the second term in (7). It can be proved easily that Gauss-Newton method has good local convergence properties when the second term is significantly small.

The efficient NLS methods, such as Levenberg-Marquardt methods and structured Quasi-Newton methods, are based on Gauss-Newton method [3]. To get the global convergence without losing good local properties, the former ones try to control the step length at each iteration by using the trust region strategy to the subproblem (5). On the other hand, the later ones reserve the first-order information $J(x_k)^T J(x_k)$ of $\nabla^2 F(x)$ and apply Quasi-Newton method to approximate the second term in (7) [4, 5].

## III. LAGRANGIAN DUAL METHOD

Consider the following problem; model of nonlinear functions that can depend on multiple parameters:

$$\min_{\substack{x \in R^{n_1} \\ y \in R^{n_2}}} \|A(y)x - b(y)\|_\infty \quad (8)$$

where, $A(y) \in R^{m \times n_1}$, $b(y) \in R^m$, ( generally $m > n_1 + n_2$ ), are nonlinear and

$$\|x\|_\infty = \max_{1 \le i \le n} |x_i|, \quad (x \in R^n).$$

This type of problems is very common and has a wide range of applications in different areas [1-8].

Generally, the problem (8) is difficult to be solved because of both the nonlinearity of $A(y)x - b(y)$ and the nondifferentiality of the infinity-norm [6]. It can be written as:

$$\min_{(x,y)} \max_i |(A(y)x - b(y))_i|, \quad i = 1, 2, ..., m \quad (9)$$

and this can be considered equivalent to the following problem in the sense of that their optimal solutions are equal

$$\min_{(x,y)} \max_i [(A(y)x - b(y))_i]^2, \quad i = 1, 2, ..., m \quad (10)$$

This is equivalent to

$$\min_{(x,y)} \max_{\substack{\lambda \in \{0,1\}^n \\ \sum_i \lambda_i = 1}} \sum_{i=1}^m \lambda_i [(A(y)x - b(y))_i]^2 \quad (11)$$

We relax it to (12)

$$\min_{(x,y)} \max_{\substack{\lambda \ge 0 \\ \sum_i \lambda_i = 1}} \sum_{i=1}^m \lambda_i [(A(y)x - b(y))_i]^2 \quad (12)$$

Actually, the optimal objective values of (11) and (12) are the same due to the fact that $\{0,1\}^n$ is the extreme points set of

$$\{\lambda : \sum_{i=1}^m \lambda_i = 1; \lambda_i \ge 0, i = 1, 2, ..., m\},$$

and any solvable linear programming problem always has a vertex solution.

The problem (12) has the following dual:

$$\max_{\substack{\lambda \ge 0 \\ \sum_i \lambda_i = 1}} \min_{(x,y)} \sum_{i=1}^m \lambda_i [(A(y)x - b(y))_i]^2 \quad (13)$$

And this problem can be solved using the subgradient method by iteratively solving its nonlinear least squares subproblems [6, 7]. The algorithm is as follows:

**Algorithm**

**Step 1** Choose the initial values $\lambda^0$ and $(x_0, y_0)$.

**Step 2** Solve the following least squares problem for fixed $\lambda^0$ with the initial solution $(x_0, y_0)$

$$\min_{(x,y)} \sum_{i=1}^{m} \lambda_i^0 [(A(y)x - b(y))_i]^2 \quad (14)$$

and obtain a local optimal solution denoted by $(x_1, y_1)$.

**Step 3** If the stop conditions satisfied, such as the variance between the current and the next obtained objective values is small enough, then stop. Otherwise, update $x_0 := x_1$, $y_0 := y_1$, $\lambda_i^0 := \lambda_i^0 + \alpha[(A(y_1)x_1 - b(y_1))_i]^2$
with $\alpha = \dfrac{1}{k + \alpha_0}$ where $k$ is the number of iterations and $\alpha_0$ is a constant; then go to Step2.

*CONCLUSIONS*

In this paper, we give a short note on NLS problem with methods which are more efficient than general unseparated ones, methods based on this scheme have the same convergence properties as the variable projection scheme theoretically and it can be applied to the algorithms for general constrained problem. Moreover, we propose a Lagrangian dual algorithm which results in solving a series of least square separable problems.

*REFERENCES*

[1] Golub G. H., and Pereyra V., "Separable nonlinear least squares: The variable projection method and its applications", Inverse Problems, Vol.19, pp. 1-26, 2002.
[2] Golub G. H., and Pereyra V., "The differentiation of pseudo-inverses and nonlinear least squares problems whose variables separate", SIAM Journal on Numerical Analysis, Vol.10: pp. 413-432, 1973.
[3] Moré J.J., "The Levenberg-Marquardt algorithm: implementation and theory", Numerical Analysis, Springer-Verlag, Vol. 630, pp. 105-116, 1978.
[4] Kaufman L., "A variable projected method for solving separable nonlinear least squares problems", BIT, Vol.15, pp. 49-57, 1975.
[5] Liu X., "An efficient unseparated scheme for separable nonlinear least squares problem", Proceedings of the eighth national conference of operations research society of China, pp. 132-137, June 30-July2, 2006.
[6] Gharibi W. and Xia Y., "A dual approach for solving nonlinear infinity-norm minimization problems with applications in separable cases", Numer. Math. J. Chinese Univ. (English Ser.), Issue 3, Vol.16, pp. 265-270, 2007.
[7] Kaufman L., and Pereyra V., "A method for nonlinear least squares problems with separable nonlinear equality constraints", SIAM Journal on Numerical Analysis. Vol. 15, pp. 12-20, 1979.
[8] Ruhe A., and Wedin P. A., "Algorithms for nonlinear least squares problems", SIAM Review Vol.22, pp. 318-337, 1980.
[9] Cheney E. W., "Introduction to approximation theory", McGraw-Hill Book. Corporation. , 1966.
[10] Weigl K. and Berthod M., "Neural networks as dynamical bases in function space", Report No.2124, INRIA Prog. Robotique, Image et Vision. Sophia Antipolis, France 1993.
[11] Weigl K., Giraudon G. and Berthod M., "Application of projection learning to the detection of urban areas in SPOT satellite images, Report No. 2143, INRIA Prog. Robotique, Image et Vision. Sophia Antipolis, France 1993.
[12] Bishop C. M., "Neural Networks for Pattern Recognition", Oxford University Press 1995.